\pdfoutput=1

\documentclass[11pt]{article}

\usepackage[utf8]{inputenc}    
\usepackage{newunicodechar}
\newunicodechar{−}{-}
\usepackage{float}

\usepackage{booktabs}  
\usepackage{colortbl}  
 
\usepackage[]{acl} 
\usepackage{graphicx}
\usepackage{times}
\usepackage{latexsym}
\usepackage{booktabs} 
\usepackage{array}
\usepackage[T1]{fontenc}
\usepackage{multirow}
\usepackage{tikz} 

\usepackage{enumitem}
\usepackage{booktabs}
\usepackage{titlesec}
\usepackage[utf8]{inputenc}
\usepackage{amsmath}
\usepackage{microtype} 
\usepackage{booktabs}
\usepackage{multirow} 

\usepackage{caption}
\usepackage{inconsolata} 
\usepackage{array} 

\newcommand\blfootnote[1]{%
  \begingroup
  \renewcommand\thefootnote{}\footnote{#1}%
  \addtocounter{footnote}{-1}%
  \endgroup
}

\title{BAGELS: Benchmarking the Automated Generation and Extraction of Limitations from Scholarly Text}
% ...

\author{\normalfont
    Ibrahim Al Azher$^{\dagger}$, 
    Miftahul Jannat Mokarrama$^{\dagger}$, 
    Zhishuai Guo$^{\dagger}$, \\
    Sagnik Ray Choudhury$^{\ddagger}$, 
    Hamed Alhoori$^{\dagger}$ \\
    $^{\dagger}$Northern Illinois University, DeKalb, IL, USA \\
    $^{\ddagger}$University of North Texas, Denton, TX, USA \\
    \texttt{\{iazher1, mmokarrama1, zguo, alhoori\}@niu.edu}, \\
    \texttt{sagnik.raychoudhury@unt.edu}
}

\begin{document}
\maketitle
\begin{abstract}


In scientific research, ``limitations'' refer to the shortcomings, constraints, or weaknesses of a study. A transparent reporting of such limitations can enhance the quality and reproducibility of research and improve public trust in science. However, authors often underreport limitations in their papers and rely on hedging strategies to meet editorial requirements at the expense of readers' clarity and confidence. This tendency, combined with the surge in scientific publications, has created a pressing need for automated approaches to extract and generate limitations from scholarly papers. To address this need, we present a full architecture for computational analysis of research limitations. Specifically, we (1) create a dataset of limitations from ACL, NeurIPS, and PeerJ papers by extracting them from the text and supplementing them with external reviews; (2) we propose methods to automatically generate limitations using a novel Retrieval Augmented Generation (RAG) technique; (3) we design a fine-grained evaluation framework for generated limitations, along with a meta-evaluation of these techniques. Code and datasets are available at: 
Code: \url{https://github.com/IbrahimAlAzhar/BAGELS_Limitation_Gen}
Dataset: \url{https://huggingface.co/datasets/IbrahimAlAzhar/limitation-generation-dataset-bagels}

\blfootnote{Accepted to the Findings of the 2025 Conference on Empirical Methods in Natural Language Processing (EMNLP 2025).}
\end{abstract}

\section{Introduction}



\begin{figure}[ht]
    \centering
    \includegraphics[width=1\linewidth, height=0.95\linewidth]{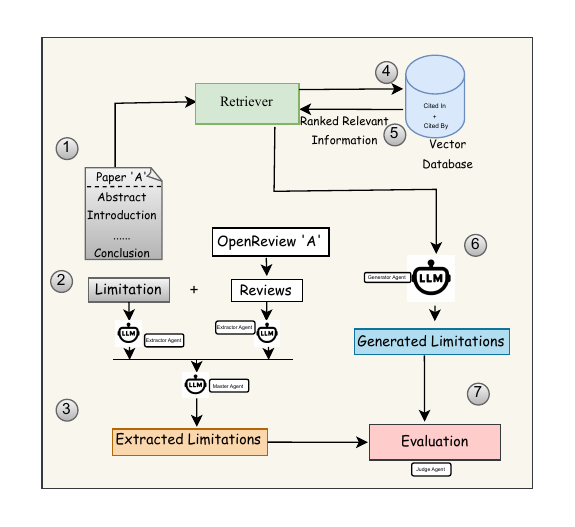}  
    \caption{System architecture for dataset creation, limitation generation, and evaluation.} 
    
    \label{fig:arch}
    \vspace{-14pt}  
\end{figure}

In scientific articles, ``limitations'' refer to the inherent shortcomings, constraints, or weaknesses of a study that may influence its results or restrict the generalizability of its findings \cite{ross2019limited}. Such limitations can arise from various aspects of the research process, including the methodology, theoretical framework, data collection, experimentation, and analysis \cite{ioannidis2007limitations}. Authors commonly acknowledge issues such as internal validity concerns, measurement errors, confounding factors, and the omission of important variables \cite{puhan2009acknowledging}. 

Openly discussing limitations is crucial. It upholds credibility and scientific integrity by demonstrating a commitment to ethical and transparent research practices \cite{bunniss2010research, chasan2014writing, annesley2010discussion,vzydvziunaite2018}. It also clarifies the scope of a study, supporting accurate interpretation, transferability, and reproducibility \cite{ioannidis2007limitations, eva2008s}. In addition, it helps researchers avoid repeating the same shortcomings \cite{escande2016limitations} while creating opportunities to refine methods and guide future research \cite{azher2025futuregen}.

Despite these benefits, researchers are often reluctant to include limitations or articulate them in detail \cite{ioannidis2007limitations, ter2013all}. Concerns about the potential impact on publication chances and career progression \cite{montori2004users} can reinforce this tendency. Even when required to acknowledge limitations, as is now common in NLP/ML research, authors sometimes resort to generic or irrelevant statements that obscure the study’s real constraints \cite{ross2019limited}. Moreover, limitations may serve as a form of \textit{hedging}, where findings are presented cautiously to avoid making definitive claims \cite{hyland1998hedging}. This practice, while safer for authors, reduces the clarity and usefulness of the research. 

Failure to disclose limitations undermines the scientific process and misleads readers, reviewers, and policymakers, preventing recognition of constrained findings and potential biases \cite{greener2018research}. Meanwhile, the volume of scientific publications has surged \cite{bornmann2021growth}. These factors highlight the need for computational methods to study research limitations. However, progress in NLP toward automatic extraction, generation, and evaluation of limitations remains limited, largely due to the lack of standardized datasets, novel methods, and robust evaluation frameworks. This study takes a step toward closing this gap.


Our contributions are as follows (see Figure \ref{fig:arch}): 
\begin{itemize}[leftmargin=*]
    \item \textbf{Dataset creation}. We build a dataset of research limitations by extracting them from papers and their reviews. By integrating author-reported and reviewer-identified limitations, this benchmark reduces self-reporting bias and provides a broader, more reliable resource for analyzing limitations and their impact on research. 
    \item \textbf{Limitation generation}. We design a novel RAG system to automatically \textit{generate} limitations, offering a way to supplement papers with high-quality, context-aware limitation statements.
    \item \textbf{Evaluation framework}. We introduce a new evaluation paradigm for generated limitations. Unlike traditional metrics (e.g., ROUGE \cite{lin2004rouge}, BLEU \cite{papineni2001bleu}, BERTScore \cite{zhang2019bertscore}, MoverScore \cite{zhao2019moverscore}), which overemphasize common terms (e.g., bias, dataset, and generalizability), our framework leverages LLMs-as-judges for fine-grained, interpretable assessments and actionable error analysis.

\end{itemize}

\section{Related Work} 
Several studies have examined how limitations are reported in papers. \citet{ioannidis2007limitations} found that only 17\% of top-tier articles mentioned limitations, with just 1\% doing so in abstracts. Similarly, \citet{puhan2012discussing} reported that 27\% of biomedical papers lacked limitations, risking overestimation of research reliability. \citet{goodman1994manuscript} noted that acknowledging limitations is often problematic in peer review. Few journals require discussing limitations \cite{ioannidis2007limitations}, which can bias reviews and weaken scientific dialogue \cite{horton2002hidden}, highlighting the need for greater transparency.

Recent work has explored computational approaches to research limitations. \citet{faizullah2024limgen} proposed an LLM-chain pipeline to summarize and refine candidate limitations. \citet{ibra2024limtopic} integrated topic modeling with LLMs to derive structured limitation themes. \citet{azher2024generating} developed a graph-augmented LLM method for generating detailed limitation statements. Other studies address the shortcomings of visualizations by generating more meaningful captions for charts and graphs \cite{al2024mitigating}. However, these studies are limited to ACL/EMNLP corpora and rely on author‐stated limitations, and use metrics such as ROUGE and BERTScore that miss finer-grained contextual alignment. 
Concurrent to our work, \citet{xu2025can} also address these gaps by introducing LIMITGEN, a benchmark that incorporates human-written peer reviews to systematically evaluate how well LLMs identify limitations. Our framework complements this effort by additionally leveraging cited papers for broader context and introducing a novel limitation-level evaluation method to preserve granularity.

Evaluating NLP outputs is essential for assessing quality, accuracy, and relevance. Traditional metrics like ROUGE and BLEU struggle with semantics, while BERTScore improves similarity but relies on references and lacks meaningful error analysis. Advances in large language models (LLMs) have opened new evaluation avenues \cite{zheng2023judging}, from zero-shot and in-context learning \cite{wei2022emergent} to specialized approaches such as GPTScore \cite{fu2023gptscore}, TIGERScore \cite{jiang2023tigerscore}, and PandaLM \cite{wang2023pandalm}. Other methods include AttrScore \cite{yue2023automatic}, which checks factual support, and SummacConv \cite{laban2022summac}, which filters low-entailment sentences. Despite their promise, LLM-based evaluations face issues such as positioning bias, where input order can shift results. We address this by randomizing order and retaining stable outputs. More broadly, our evaluation advances beyond prior work by combining granularity-aware scoring, topic-level agreement, and LLMs-as-judges.

Taken together, prior research shows both the need and the opportunity for a more systematic treatment of research limitations. Building on these insights, our work unifies dataset construction, limitation generation, and evaluation into a single framework, laying the foundation for more transparent and reproducible analysis of limitations.

\section{Limitation Extraction \& Evaluation}
 
\subsection{Dataset of Extracted Limitations}
\label{sec:extraction}
\textbf{Granularity. }
A key challenge in building a dataset of research limitations is defining the appropriate level of granularity. Should a limitation be captured as a single phrase, a full sentence, or an entire paragraph? We define a \textbf{limitation} as a \textit{sequence of sentences}, as individual sentences often do not encapsulate multiple limitations. In contrast, a single limitation can extend across multiple sentences, sometimes forming a complete paragraph.

\textbf{Extraction Sources. } 
Two primary sources form the basis of our dataset: (1) limitations explicitly acknowledged by authors, and (2) those highlighted through peer-review commentary. Although author-reported limitations often provide well-structured insights, previous research indicates that such limitations may be underreported or carefully hedged. To address this gap, we incorporate review comments, where reviewers often highlight additional constraints or weaknesses not mentioned by the authors.

Our dataset includes papers from major NLP and ML conferences, including  ACL\footnote{https://aclrollingreview.org/cfp} and  NeurIPS\footnote{https://neurips.cc/public/guides/PaperChecklist}, as well as biomedical research from PeerJ \footnote{https://peerj.com/benefits/indexing-and-impact-factor/}. We collect 6,932 NeurIPS papers (2021-2022), 5,739 ACL papers (2023-2024), and 1000 papers from PeerJ. In addition, we integrate OpenReview \footnote{\url{https://openreview.net/}} comments for 2,802 papers from NeurIPS. All of the PeerJ papers contain self-reported limitations alongside other sections and peer review comments. 
For each paper, we use LLM to extract and get an average of $8$ limitations from a paper and $10$ from their reviews. 


\subsection{Extraction Process} 
We extract spans (blocks) of text from papers or review comments, and then refine them with LLMs, as opposed to passing in the entire paper to an LLM. This strikes a balance between accuracy and LLM usage cost.

\textbf{1. Limitation \textit{Span} Extraction:} 
This step extracts blocks of text from the papers that correspond to limitations. We consider both explicit and implicit limitation statements:

\textbf{a. Explicit limitations.} These appear in a dedicated limitations section or subsection. We identify them using the AllenAI Science Parse tool \footnote{https://github.com/allenai/science-parse}, which segments papers into a structured JSON format, allowing for direct and reliable extraction of these dedicated sections. For peer review content in NeurIPS papers, we used Selenium to scrape the main review field from OpenReview, which typically includes both strengths and weaknesses of a paper.

\textbf{b. Implicit limitations.} These are embedded in broader sections such as discussion or conclusion. To identify them, we apply a Python regex script that searches for keywords such as \textit{limitation(s)}, or \textit{shortcoming(s)}. To improve precision, we exclude sections where limitations are rarely discussed (e.g., abstract, introduction, related work). Our script begins extraction when a limitation-related keyword is detected and continues until a terminal section marker is reached; extraction stops at terms such as acknowledgements, grant, future work, discussion, conclusion, or appendix. Although this process is effective, the regex approach for implicit limitations can occasionally capture irrelevant sentences, introducing noise into the results.

\textbf{2. Refinement via LLM:} 
To improve precision, we use an LLM to filter meaningful limitations from the tool-extracted ones (from both papers and review) by removing noisy sentences. Importantly, we strictly instruct the LLMs to extract limitation statements without paraphrasing, altering, or generating new content, and producing them as a structured sequence of sentences, denoted by 
\(L_i = \{l_{i1}, l_{i2}, \dots, l_{ix}\}\). 
To incorporate broader perspectives from peer reviews, we first aggregate comments from multiple review responses into a single consolidated text. We then prompt the LLMs (Figure~\ref{fig:prompt_topic_generation_related_text}, appendix) to segment this text and identify distinct limitation statements by reviewers, with the latter being denoted as \(R_i = \{r_{i1}, r_{i2}, \dots, r_{ix}\}\).

Following this extraction, a master LLM is tasked with merging the author-reported limitations $L_i$ and the reviewer-identified limitations $R_i$ of input paper $P_i$. The model is explicitly instructed to merge only those limitation statements that were identical or semantically equivalent across both the author-mentioned limitations and the peer review. As before, the model is restricted from changing, rephrasing, or reordering any sentences during the merge process, and we get final \emph{Ground truth extracted limitations} \(G_i = \{g_{i1}, g_{i2}, \dots, g_{ix}\}\). Finally, we evaluate the quality of these extracted and merged limitations through a user study described in \S~\ref{sec:humeval}. We use GPT 4o-mini as both the extractor and master LLM (Examples of limitations extraction by LLM from NeurIPS, ACL, and OpenReview are provided in the Appendix in Figure \ref{fig:llm_refined_limitations_in_neurips}, \ref{fig:llm_refined_limitations_in_neurips1}, and \ref{fig:llm_refined_limitations_in_neurips3}, respectively).

\subsection{Limitation Extraction Evaluation} 
\label{sec:humeval}  

\textbf{Are the limitations extracted or generated?} The first goal in the evaluation process is to check if the LLM extracted limitations are \textbf{grounded in the text}, i.e., they only come from the input (papers/reviews) and not from the LLMs' parametric knowledge or hallucinations. For this, we employ three annotators (separate from this paper's authors)\footnote{CS graduate students with research experience in NLP and AI}. 

The first ground truth consists of only author-mentioned limitations. We choose a sample of 100 limitations from ACL, NeurIPS, and PeerJ, and for each, we show them the source and ask a Yes/No question, whether they thought the LLM \textit{extracted} the limitation from the source without generating text. Each annotator answer positively in $> 90\%$ of cases (avg $\pm$ std= $95 \pm 2.45\%$) (Table \ref{table:eval-gpt-with-human-annotator}).

\begin{table}[htbp]
  \centering
  \footnotesize
  \begin{tabular}{l c c c c c}
    \toprule
    \textbf{Model}  & \textbf{Role}  & \textbf{Sample} & \textbf{U1} & \textbf{U2} & \textbf{U3} \\  
    \midrule
    GPT 4o-mini & Extractor & 100 & 92 & 95 & 98 \\  
    \bottomrule
  \end{tabular}
  \caption{Evaluating LLM as an extractor role with human annotator (U).}
  \label{table:eval-gpt-with-human-annotator} 
  \vspace{-5pt}
\end{table}

In the second evaluation, two annotators manually verified the extracted limitations from 1000 papers from NeurIPS and PeerJ \textit{and their reviews}. The annotators assessed whether 1) each LLM-extracted author mentioned limitation was grounded in the source paper, (2) each extracted limitation from the peer review was also grounded in the review, and (3) the merged set (limitation + review) included only truly overlapping or matching limitations between the two sources. Their analysis confirmed that all extracted limitations were faithfully sourced, with no instances of hallucinated, noisy, or newly generated content. We also computed the performance of the Llama3 70B for this extraction task, and the result was unsatisfactory.

\textbf{The quality of the extraction.} The SMEs from the last step annotated 500 ACL papers and 100 NeurIPS papers: one annotator extracted limitations (taking the full section when explicit, or selecting limitation-related sentences when implicit), and two others verified the results. We then compared the tool-based (GPT-4o mini) extractions against this gold standard. Notably, the human-extracted (gold) limitations were not segmented; therefore, we combined the LLM-extracted limitations and compared them with the gold ones using cosine similarity, precision, recall, F1, and fuzzy matching\footnote{These strings are tokenized.} (Table \ref{table:eval-with-human-curated-acl-neurips}): ACL achieved a strong F1 of 85.69, likely aided by more frequent explicit limitation sections. NeurIPS yielded a moderate F1 of 72.42, reflecting the more scattered, implicit presentation of limitations where LLM should be utilized to remove noisy information.

\begin{table}[!htbp]
  \centering
  \footnotesize
  
  \begin{tabular}{l c c c c c}
    \toprule
    \textbf{Dataset} & \textbf{CS} & \textbf{P} & \textbf{R} & \textbf{F1} & \textbf{Fuzzy} \\  
    \midrule
    ACL & 89.38 & 89.63 & 84.93 & 85.69 & 91.18 \\ 
    Neurips & 78.08 & 68.76 & 84.13 & 72.42 & 70.26 \\ 
    \bottomrule
  \end{tabular}
  \caption{ Performance between Human Extracted Limitations vs Tool Extracted Limitations in Cosine Similarity (CS), Precision (P), Recall (R), F1 score (F1), and Fuzzy matching}
  \label{table:eval-with-human-curated-acl-neurips} 
  \vspace{-5pt}
\end{table}

\subsection{Dataset Applications}
The resulting dataset is publicly available\footnote{\url{https://huggingface.co/datasets/IbrahimAlAzhar/limitation-generation-dataset-bagels}} and can be used as a benchmark for evaluating automated limitation extraction and generation methods (\S\ref{sec:eval}). Beyond this, the extracted limitations can be examined and organized into a taxonomy of limitations in ML and NLP, offering a more structured understanding of common research challenges. By integrating this taxonomy into citation networks, we can introduce the concept of a \textit{Limitation Multigraph}, enabling scientometric analyses into whether certain limitations shape the direction of subsequent research or, alternatively, tend to be overlooked. These avenues present new opportunities to study how the reporting (or the lack thereof) of limitations affects the broader scientific discourse, a topic we plan to explore in future work.

\section{Limitation Generation} 
Most research papers either do not explicitly mention limitations or underreport them, even when a dedicated section is provided. We compare two systems' ability to generate limitations from research papers: (a) vanilla LLM and (b) RAG. Note that the generators don't have access to the text from where the limitations are extracted, e.g., limitation sections of the papers, paragraphs identified as limitations, or paper reviews; otherwise, the task would be trivial. To improve computational efficiency, we use the three most important sections of a paper as input to the generators rather than the full text. The importance score is computed by the cosine similarity of a section and a reference limitation embedding (see Table \ref{tab:cosine-sim}, Appendix).

\textbf{Vanilla LLM. } 
In the \emph{vanilla} LLM setup, when the input exceeds the context window, it is divided into chunks \(\{P_i^\prime\}\), and limitations are generated for each chunk \cite{d2024marg}. The LLM is then also asked to aggregate these chunk-specific outputs into a cohesive, meaningful final set of limitations.

\textbf{RAG Integration. }
A paper \(P_i\) can be used independently to generate limitations, but this approach risks overlooking valuable insights from other, potentially \emph{related} papers. In particular, even when a paper lacks an explicit limitations section, other papers with similar methodologies or datasets may discuss relevant shortcomings. For example, a paper can use SVM and not explicitly mention the modeling assumptions, whereas a related paper possibly will. Moreover, certain findings may be implicitly contradicted by subsequent research.
To address this issue, we employ a RAG framework, which allows the system to draw context from multiple papers rather than relying only on \(P_i\).

There can be multiple notions of relatedness; we compare between two: a) relatedness induced by the citation network of $P_i$, and b) textual similarity between $P_i$ and other papers. For the citation network, we use both the P\textsubscript{cited-by, i}: papers citing $P_i$, and P\textsubscript{cited-in, i}: papers that $P_i$ cites. We parse the reference section of $P_i$ to extract the DOI and title of each ``cited in'' work. The ``cited by'' DOI and titles are collected from the OpenAlex API \footnote{https://openalex.org/}. We query the Semantic Scholar API \footnote{\url{https://www.semanticscholar.org/product/api}} with $P_i$'s title to get the DOIs for \textbf{top 5} most semantically close papers. These DOIs and titles are cross-referenced with arXiv metadata, and the full texts of the matched papers are downloaded and parsed with the Science Parse tool.


For each paper $P_i$, we build separate RAG indices with a) P\textsubscript{cited-by, i} b) P\textsubscript{cited-in, i}, c) semantically close papers, and their combinations, where papers are split into chunks by section to preserve detail. We combine the strengths of both keyword-based (BM-25) and semantic (FAISS) search by assigning a 50\% weight to the scores from each retriever. We use a \textit{LLM-based reranker}, where we retrieve 20 chunks, and pass these chunks to a GPT 4o-mini model along with the original input paper. The model is prompted to score the relevance of each chunk on a scale of 1 to 10. Only the chunks that receive a relevance score of 8 or higher are ultimately selected. We compare this method with the simple baseline of just using the retrieved chunks in \S \ref{sec:ablation-study}.

\begin{figure}[ht]
    \centering
    \includegraphics[width=1.\linewidth]{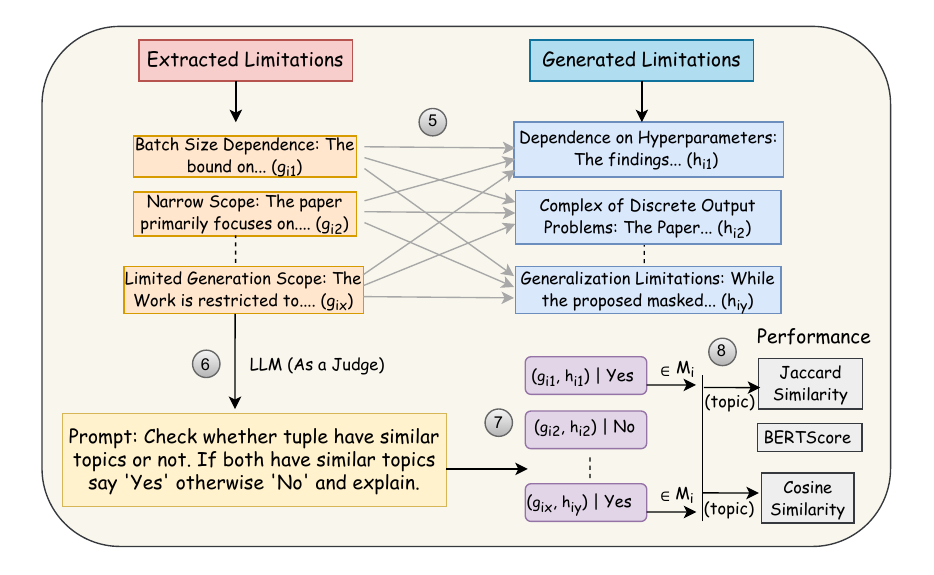}  
    \caption{Evaluation of generated limitations.}
    \label{fig:arch_eval}
    \vspace{-13pt}  
\end{figure}

\section{Evaluation of Generated Limitations}
\label{sec:eval}
We want to evaluate the quality of the generated limitations by comparing them with the \emph{extracted} ones. Functionally, both the ground-truth and the predictions are a set of text segments. NLP metrics like BERTScore, ROUGE, and cosine similarity can yield surface overlaps, providing high scores even when the generated limitations are not appropriate, too generic, or imprecise. A possible alternative is to use a holistic LLM-as-Judge approach, where the generated/ground-truth limitations are merged into single text blocks and then compared. This lacks the point-level granularity needed for fine-grained analysis. We address both these problems by introducing the \emph{PointWise} (\textbf{PW}) evaluation framework (Figure \ref{fig:arch_eval}).


\paragraph{Problem Setup.}
Suppose we have a set of papers \(P = \{P_1, P_2, \dots, P_n\}\). For each paper \(P_i\), we assume access to: \emph{Ground truth limitations} \(G_i = \{g_{i1}, g_{i2}, \dots, g_{ix}\}\), where \(x\) is the number of ground truth limitations we extracted or annotated for \(P_i\). And \emph{LLM-generated limitations} \(H_i = \{h_{i1}, h_{i2}, \dots, h_{iy}\}\), where \(y\) is the number of limitations produced by the LLM for \(P_i\).
Our goal is to measure (1) how many ground truth limitations the LLM correctly reproduces (\emph{coverage}) and (2) how well each matched pair of limitations aligns in content and focus (\emph{performance}).



\subsection{Coverage}
\paragraph{A. Pairwise Matching.}

To quantify coverage, we first create all possible pairs of limitations between the sets \(G_i\) and \(H_i\). Let
\[
S_i = \{(g_{ik},\, h_{il}) \mid 1 \le k \le x,\, 1 \le l \le y\}.
\]

Hence, \(|S_i| = x \times y\). We then use an LLM \emph{as a judge} \cite{zheng2023judging} to decide if a ground truth limitation \(g_{ik}\) and a generated limitation \(h_{il}\) are similar in content or topic (Figure \ref{fig:llm_as_a_judge_for_sub_limitations}, Appendix):
\[
J(g_{ik}, h_{il}) \;=\;
\begin{cases}
1, & \text{if } g_{ik} \text{ and } h_{il} \text{ are similar},\\
0, & \text{otherwise}.
\end{cases}
\]

We collect all \emph{matched} pairs into a set
\[
M_i \;=\; \{(g_{ik},\, h_{il}) \mid J(g_{ik},\, h_{il}) = 1\},
\]
and let \(\lvert M_i\rvert = z_i\) be the number of matched pairs for paper \(P_i\).




\paragraph{B. Coverage of Ground Truth Limitations.}
We define \(C_{Gi}(g_{ik}) = 1\) if the ground truth limitation \(g_{ik}\) appears in \emph{at least one} matched pair in \(M_i\), and 0 otherwise:
\[
C_{Gi}(g_{ik}) \;=\;
\begin{cases}
1, & \exists\,h_{il}\text{ such that }(g_{ik},\,h_{il}) \in M_i,\\
0, & \text{otherwise}.
\end{cases}
\]
The \emph{coverage of ground truth limitations} for paper \(P_i\) is
\[
A_{Gi} \;=\; \frac{1}{x} \sum_{k=1}^{x} C_{Gi}(g_{ik}).
\]
In other words, \(A_{Gi}\) measures the fraction of ground truth limitations in \(P_i\) that are matched with at least one LLM-generated limitation. 



\paragraph{C. Coverage of LLM-Generated Limitations.}
Similarly, we define \(C_{Hi}(h_{il}) = 1\) if a generated limitation \(h_{il}\) appears in \emph{at least one} matched pair in \(M_i\), and 0 otherwise:
\[
C_{Hi}(h_{il}) \;=\;
\begin{cases}
1, & \exists\,g_{ik}\text{ such that }(g_{ik},\,h_{il}) \in M_i,\\
0, & \text{otherwise}.
\end{cases}
\]
The \emph{coverage of LLM-generated limitations} for paper \(P_i\) is
\[
A_{Hi} \;=\; \frac{1}{y} \sum_{l=1}^{y} C_{Hi}(h_{il}).
\]
We aggregate these coverage values across all pa-
pers by taking their means:
\[
A_{G} \;=\; \frac{1}{n} \sum_{i=1}^{n} A_{Gi}, 
\quad
A_{H} \;=\; \frac{1}{n} \sum_{i=1}^{n} A_{Hi}.
\]
Finally, we display $A_G$ and $A_H$ as percentage. 




\paragraph{D. Precision, Recall, and F\textsubscript{1}.}
We also compute overall precision, recall, and F\textsubscript{1} scores. For each paper \(P_i\):


\begin{align*}
\text{TP}_i &= |M_i|, \\
\text{FP}_i &= x - \sum_{k=1}^{x} C_{Gi}(g_{ik}), \\
\text{FN}_i &= y - \sum_{l=1}^{y} C_{Hi}(h_{il}).
\end{align*}

Here, \(\text{TP}_i\) (\emph{true positives}) is the total number of matched pairs; \(\text{FP}_i\) (\emph{false positives}) is the number of ground truth limitations not matched by any LLM-generated limitation; \(\text{FN}_i\) (\emph{false negatives}) is the number of LLM-generated limitations unmatched by any ground truth limitation. True negative  $(TN_{i})$ is not applicable in this case, as we do not have a defined \textit{negative} class. If there is one ground truth limitation $g_{ik}$ that matches with multiple LLM-generated limitations (and vice versa), True Positive (TP) counts as one. (Details in Appendix \ref{sec:appendix}.1)

\subsection{Performance}
After identifying matched pairs  \((g_{ik}, h_{il}) \in M_i\), we score each pair's quality using (i) text-based metrics: ROUGE-L, BERTScore, and cosine similarity, and (ii) keyword overlap (Jaccard Similarity). Finally, the per-pair scores are averaged. Unmatched items are excluded, as our goal is to quantify similarity within aligned pairs rather than coverage (details in Appendix \ref{sec:appendix}.2).

\section{Experimental Setup for Generation}
\label{sec:exp-setup} 

We use three LLMs (GPT-3.5, GPT-4o-mini, and Llama 3.1 8B \footnote{https://huggingface.co/meta-llama/Llama-3.1-8B}) in a zero-shot setup for both the vanilla generation and RAG. The GPT models are accessed through APIs, and the LLama models are locally deployed with Ollama.
For the vanilla generation, we also fine-tune three sequence-to-sequence models, T5 (512-token window) \cite{raffel2020exploring}, BART (1024 tokens), and Pegasus (1024 tokens) \cite{zhang2020pegasus} on a 70 / 30 train–test split. All models were trained for 3 epochs with a learning rate of $5\times10^{-5}$, weight decay 0.01, 300 warmup steps, and batch sizes of 4 (train) and 8 (eval), with early stopping; inputs longer than 512 tokens were truncated.
For RAG, the vector database is built with llama-index \footnote{https://www.llamaindex.ai/}, and the OpenAI text-embedding-ada-002 embedding model for encoding the source and query documents.

\section{Experiments and Results}


\textbf{Evaluation of LLM as Aligner.}
The PointWise evaluation protocol above uses an LLM to determine whether a generated limitation matches or \textit{aligns with} a ground truth one. We evaluate GPT 4o-mini's reliability in this task. A set of 100 positive (as per the model prediction) and 100 negative instances is annotated independently by three human evaluators. The human annotators have a Cohen's $\kappa$ score of $>=$ 95\% (Table \ref{table:human-v-model-matching}, Appendix), which shows that the task is largely unambiguous. Cohen's $\kappa$ between human annotators and model (GPT 4o mini) prediction is 90-95\%, showing exceptional agreement. In comparison, Llama-3.1 400B shows poor agreement with the human judges (76\%-81\%), so in subsequent evaluations, we use GPT-4o mini as the aligner. See Table \ref{tab:paper_gt_vs_gen_lim_vs_human} in the appendix, for an example alignment.

\subsection{Limitation Generation Evaluation} 
The ground truth contains papers that have a) only self-reported limitations and b) limitations coming from both self-reports and reviews. 

\subsection{Author-Mentioned Limitation}
We evaluate the model's ability to generate self-reported limitations on the ACL part of the dataset, as these papers a) have explicit limitation sections, and b) do not have open-access reviews. The results are presented in Table \ref{tab:performance-coverage-accuracy}.   

\textbf{Vanilla LLMs and fine-tuned models. } 
Zero-shot models outperform trained models in almost all metrics, with GPT-3.5 achieving the best results in coverage metrics, and LLama 3 achieving the best in performance metrics. Surprisingly, GPT 4o-mini has a significantly worse performance than other zero-shot models. However, the performance metrics are based on n-gram overlap and embedding measures (e.g., ROUGE, BLEU, BERTScore, cosine similarity) that primarily capture surface overlap or shallow semantics and can miss factual correctness and completeness. Therefore, we prioritize coverage-based metrics, C\textsubscript{GT}, C\textsubscript{LLM}, and F1, and report NLP metrics as secondary diagnostics. 

\textbf{RAG. } 
Since GPT-3.5 performs the best in the coverage metrics, we utilize it in a RAG setup, where the index consists of ``cited-in'' and ``cited-by'' papers. This improves the performance metrics, but comes at a cost of coverage metrics. To understand whether this reduction is caused by the RAG setup or the model, we include GPT 4o-mini in the same RAG setup, which shows a significant improvement in all metrics.


\begin{table*}[!htbp]
  \centering
  \footnotesize
    \begin{tabular}{c c c c c c c c c c c}
    \toprule
     \textbf{Model} & \textbf{Model type}  & \textbf{R-L} & \textbf{BS}  & \textbf{JS} & \textbf{CS} & \textbf{C\textsubscript{GT}} & \textbf{C\textsubscript{LLM}} & \textbf{Prec.} & \textbf{Recall} & \textbf{F1} \\ 
    \midrule
    T5 & fine-tuned & 19.92 & 87.81 & 10.82 & 31.79 & 35.48 & 29.59 & 0.29 & 0.31 & 0.30\\ 
    BART & fine-tuned & 19.43 & 87.67  & 10.68 & 31.91 & 33.71 & 30.10 & 0.30 & 0.31 & 0.31 \\  
    Pegasus & fine-tuned & 20.15 & 87.66 & 10.71 & 33.39 & 29.28 & 25.27 & 0.25 & 0.26 & 0.26 \\
    \midrule
    Llama 3 & zero-shot & 25.66 & 88.30 & 14.69 & 40.4  & 61.38 & 39.04 & 0.39 & 0.50 & 0.44\\ 
    GPT-3.5 & zero-shot & 24.24 & 87.08 & 14.65 & 43.12  & \textbf{76.62} & \textbf{46.65}  & 0.47 & \textbf{0.67} & \textbf{0.55}  \\
    GPT 4o-mini & zero-shot & 16.57 & 86.02 & 8.70 & 32.29 & 57.65 & 19.76 & 0.20 & 0.31 & 0.24 \\ 
    \midrule
    GPT-3.5 + RAG & zero-shot + RAG & \textbf{30.21} & \textbf{90.88} & \textbf{19.47} & \textbf{45.37} & 39.99 & 44.66 & 0.42 & 0.40 & 0.41 \\ 
    GPT 4o-mini + RAG & zero-shot + RAG & 23.17 & 87.33 & 12.99 & 39.29 & 67.13 & 45.67 & \textbf{0.57} & 0.45 & 0.51 \\ 
    
    \bottomrule
  \end{tabular} 
    \caption{Results of models in ``Coverage'' (Coverage of Ground Truth Limitation (C\textsubscript{GT}), LLM Generated Limitation (C\textsubscript{LLM}), Precision, Recall, and F1-score) and ``performance'' metrics -- \textbf{R}ouge-x, BLEU, BertScore (BS), Jaccard (JS) and Cosine (CS) similarity on the \textbf{ACL} dataset. In all metrics, a higher score denotes a better performance.}
  \label{tab:performance-coverage-accuracy}
\end{table*} 

\subsection{Self-reported \& Peer-review Limitation} 
The ground truth here consists of author-stated limitations and peer-review limitations extracted from NeurIPS papers. We hypothesize that the RAG approaches should be beneficial for this dataset, as the reviewers are more likely to point out limitations from external sources, such as cited in/by or semantically similar papers. Therefore, we use this dataset to compare different RAG approaches with GPT 4o-mini as the baseline LLM, as the previous experiments (Table \ref{tab:performance-coverage-accuracy}) suggest that it has the highest propensity of improvement with RAG. 

\begin{table*}[ht]
  \centering
  \footnotesize 
  \begin{tabular}{c c c c c c c c}
    \toprule 
    \textbf{RAG Index}  & \textbf{C\_GT} & \textbf{C\_LLM} & \textbf{F1} & \textbf{R-L} & \textbf{BS} & \textbf{CS} & \textbf{JS} \\  
    \midrule
Not applicable & 67.34 & 63.81 & 0.65 & 15.30 & 86.66 & 33.43 & 8.69 \\
100 Random Papers & 60.31 & 48.87 & 0.52 & \textbf{16.39} & \textbf{86.88} & 32.34 & \textbf{8.99} \\
Cited In & \textbf{69.27} & 62.59 & 0.65 & 14.07 & 86.37 & 33.23 & 8.05 \\
Cited By & 68.45 & 64.01 & 0.65 & 14.49 & 86.34 & \textbf{34.09} & 8.36 \\
Cited In + Cited By & 68.84 & \textbf{64.87} & \textbf{0.67} & 14.35 & 86.35 & 33.94 & 8.28 \\
Cited In + By + Semantically Similar 5 Papers & 68.02 & 63.38 & 0.65 & 14.59 & 86.39 & 33.72 & 8.38 \\
   \bottomrule
  \end{tabular} 
  \caption{ Coverage evaluation of multiple types of RAG vector database settings in \textbf{NeurIPS 21-22 dataset} with GPT 4o-mini as the base LLM.}
  \label{tab:coverage-gpt4o-vs-diff-rag-settings}
\end{table*}

Table \ref{tab:coverage-gpt4o-vs-diff-rag-settings} presents the performances with different RAG indices, with the first row representing vanilla LLM (no RAG). When the index is built with 100 random papers, the F1 score drops (-0.13) compared to the zero-shot approach. A combination of ``cited in'' and ``cited by'' papers achieves the highest F1 score of 0.67 -- an increase of 0.02 over the baseline. However, when we further add the top five semantically related papers retrieved via the Semantic Scholar API, the F1 score reduces (-0.02), indicating that including loosely related content can introduce noise and reduce overall precision. 


However, the performance metrics present a somewhat different story. Semantically related sentences from 100 randomly selected papers yield the highest scores across multiple metrics, including ROUGE-L, BERTScore, BLEU, and Jaccard similarity. We believe this is due to the vector database containing diverse texts, which are not semantically or n-gram overlapping with the ground truth. This perhaps also shows the brittleness of performance metrics for evaluating the generation quality of limitations.   

\subsection{Ablation Study}
\label{sec:ablation-study}

\textbf{a. Size of the input text:} We investigate the effect of the length of the input text on the generator models with an ablation study (Table \ref{tab:performance-of-lim-with-openreview}). We use a) GPT-4o mini + RAG and b) Llama-3.1-8B, as these are the best-performing systems in the RAG and vanilla LLM setups, considering the author-mentioned limitations, the review-mentioned ones, and their combinations. When using GPT-4o-mini with RAG, expanding the context from the top-3 sections to all available sections generally increased pointwise scores for the author-written ground truth: C\textsubscript{GT} (+0.94), C\textsubscript{LLM} (+0.82), and F1 (+0.02). A similar trend was observed for the reviewer-suggested and combined ground truths, with the exception of a slight dip in the C\textsubscript{GT} score for the combined (Auth + Rev) case. By contrast, most NLP-based metrics (e.g., ROUGE, BERTScore, and cosine) slightly decreased with all-section inputs. Taken together, this indicates that using only the top-3 sections is a cost-effective alternative in this setup: minor drops in pointwise metrics, small gains (or less drop) in NLP metrics, and no large performance loss overall. In Llama-3.1-8B, however, we observe the opposite trend. Moving to all sections produces a large F1 gain (+0.13) for the combined ground truth and improves most NLP-based metrics. This suggests the smaller Llama-3.1-8B benefits from the full-paper context to generate higher-quality limitations, whereas truncating to the top-3 sections leaves it under-informed.

\textbf{b. Retriever Method:}
On the NeurIPS dataset, we evaluate our LLM re-ranker against a vanilla retriever baseline, both operating within a RAG framework with GPT-4o mini generator (Table \ref{tab:vanila-vs-llm-reranker}). While the baseline simply retrieves the top 3 chunks using a FAISS+BM25 search, our method re-ranks the top 20 chunks, leading to substantial gains in C\textsubscript{GT} (+28.5), C\textsubscript{LLM} (+15.53), and the F1 score (+0.24).

\begin{table}[ht]
  \centering
  \footnotesize
  \begin{tabular}{c c c c c}
    \toprule
    \multirow{2}{*}{\textbf{Metric}} & \multicolumn{1}{c}{\textbf{Input}} & \multicolumn{3}{c}{\textbf{Ground Truth}} \\ 
    \cmidrule(lr){2-2} \cmidrule(lr){3-5}
    & \textbf{Sec} & \textbf{Auth} & \textbf{Rev.} & \textbf{Auth + Rev} \\ 
    \midrule
    \multicolumn{3}{l}{\textbf{GPT 4o-mini + RAG}} \\
    \midrule
    R-L & 3 & 16.93 & 11.92 & 14.44   \\
     & All & 16.73 (\scalebox{0.5}{$\downarrow$}) & 12.12 (\scalebox{0.5}{$\uparrow$}) & 14.35 (\scalebox{0.5}{$\downarrow$})  \\
     \midrule
    BS & 3 & 87.27 & 86.26 & 86.43   \\
     & All & 87.15 (\scalebox{0.5}{$\downarrow$}) & 86.21 (\scalebox{0.5}{$\downarrow$}) & 86.35 (\scalebox{0.5}{$\downarrow$}) \\
     \midrule
    CS & 3 & 36.03 & 29.16 & 33.88  \\    
     & All & 35.52 (\scalebox{0.5}{$\downarrow$}) & 29.81 (\scalebox{0.5}{$\uparrow$}) & 33.94 (\scalebox{0.5}{$\uparrow$}) \\ 
     \midrule
    C\textsubscript{GT} & 3 & 82.60 & 61.19 & 69.78  \\
     & All & 83.54 (\scalebox{0.5}{$\uparrow$}) & 61.93 (\scalebox{0.5}{$\uparrow$}) & 68.84 (\scalebox{0.5}{$\downarrow$})  \\
     \midrule
    C\textsubscript{LLM} & 3 & 29.83 & 62.69 & 61.59  \\ 
     & All & 30.65 (\scalebox{0.5}{$\uparrow$}) & 63.97 (\scalebox{0.5}{$\uparrow$}) & 64.87 (\scalebox{0.5}{$\uparrow$})  \\ 
     \midrule
    F1 & 3 & 0.40 & 0.62 & 0.64 \\ 
     & All & 0.42 (\scalebox{0.5}{$\uparrow$}) & 0.63 (\scalebox{0.5}{$\uparrow$}) & 0.67 (\scalebox{0.5}{$\uparrow$})  \\     
    \midrule
    \multicolumn{3}{l}{\textbf{Llama 3.1 8B}} \\
    \midrule
    R-L & 3 & 17.67 & 12.30 & 15.03  \\
    & All & 17.76 (\scalebox{0.5}{$\uparrow$}) & 13.75 (\scalebox{0.5}{$\uparrow$}) & 14.92 (\scalebox{0.5}{$\downarrow$})  \\ 
    \midrule
    BS & 3 & 87.34 & 86.53 & 86.65 \\
     & All & 87.54 (\scalebox{0.5}{$\uparrow$}) & 87.23 (\scalebox{0.5}{$\uparrow$}) & 87.06 (\scalebox{0.5}{$\uparrow$}) \\  
    \midrule
    CS & 3 & 31.82 & 25.16 & 30.72  \\
    & All & 31.99 (\scalebox{0.5}{$\uparrow$}) & 27.47 (\scalebox{0.5}{$\uparrow$}) & 29.99 (\scalebox{0.5}{$\downarrow$}) \\ 
    \midrule
    C\textsubscript{GT} & 3 & 63.52 & 42.79 & 44.74  \\
     & All & 64.47 (\scalebox{0.5}{$\uparrow$}) & 57.26 (\scalebox{0.5}{$\uparrow$}) & 62.04 (\scalebox{0.5}{$\uparrow$})  \\
     \midrule
    C\textsubscript{LLM} & 3 & 33.32 & 44.67 & 48.93 \\ 
    & All & 27.75 (\scalebox{0.5}{$\downarrow$}) & 55.89 (\scalebox{0.5}{$\uparrow$}) & 58.86 (\scalebox{0.5}{$\uparrow$})  \\ 
    \midrule
    F1 & 3 & 0.39 & 0.43 & 0.46 \\ 
    & All & 0.34 (\scalebox{0.5}{$\downarrow$}) & 0.56 (\scalebox{0.5}{$\uparrow$}) & 0.59 (\scalebox{0.5}{$\uparrow$}) \\ 
    \bottomrule
  \end{tabular} 
    \caption{Ablation study with GPT 4o-mini + RAG and Llama 3.1 8B results in ``coverage'' (Coverage of Ground Truth Limitation (C\textsubscript{GT}), LLM Generated Limitation (C\textsubscript{LLM}), F1-score) and ``performance'' (\textbf{R}ouge-x, \textbf{B}ert\textbf{S}core (BS), and Cosine (CS) similarity in the \textbf{NeurIPS data}.}
  \label{tab:performance-of-lim-with-openreview} 
\end{table}

\begin{table}[ht]
  \centering
  \footnotesize 
  \begin{tabular}{p{1.65cm} p{1.85cm} p{0.79cm} p{0.85cm} p{0.7cm}} 
    \toprule 
    \textbf{Model}  & \textbf{VD}  & \textbf{C\_GT} & \textbf{C\_LLM} & \textbf{F1} \\ 
    \midrule 
GPT 4o-mini  & Vanila k =3 & 40.34 & 49.34 &  0.43 \\  
GPT 4o-mini & LLM re-ranker & \textbf{68.84} & \textbf{64.87} & \textbf{0.67} \\ 
   \bottomrule
  \end{tabular} 
  \caption{Performance between different retriever approaches in VD (Vector Database) in RAG (vanilla RAG (considering top 3 chunks) vs LLM re-ranker)}
  \label{tab:vanila-vs-llm-reranker} 
\end{table}

Our findings demonstrate that a multi-faceted approach, combining curated external data with targeted retrieval, significantly enhances the generation of scientific limitations. This is especially evident when we use limitations extracted from reviews in the ground truth, as the use of ``cited in'' and ``cited by'' papers in the RAG index achieves the highest F1 score. We also observe that the length of the input to the generator model has a different effect in the vanilla LLM and RAG setup. It might be beneficial to use full paper texts for smaller models, but larger models in RAG setups can perform reasonably well with the most important parts of a paper. 
\section{Conclusion}
We present a new approach for automatically extracting, generating, and evaluating limitations in scientific articles. Our method explores incorporating cited works, accommodating top sections of the entire paper, and integrating review feedback to capture perspectives beyond those of the original authors. To evaluate the effectiveness of our system, we introduce a granular text evaluation framework that breaks down limitations into more minor points and employs LLMs as a Judge for assessing alignment. Human review validates our extraction and LLM-as-Judge pipeline, showing strong agreement with expert judgments.



\section*{Limitations}
In this work, we focused on venues in natural language processing (ACL papers from 2023-2024) and machine learning (NeurIPS papers 2021-2022), and Biology domain papers from PeerJ, which ensures high relevance and quality but insufficient for broader generalizability. While this scope allows us to benchmark the performance of LLMs in extracting limitations from well-structured scientific texts, we acknowledge that the findings may not generalize to papers from other fields, such as social sciences, physics, chemistry, or mathematics where writing conventions and limitation styles may differ. 

Due to high API costs, we did not experiment with GPT-4 or GPT-4o; instead, we opted for GPT-4o Mini as a cost-effective alternative. While we incorporated OpenReview comments for NeurIPS papers, we could not find them for ACL papers. Furthermore, we relied on GPT-4o Mini as the evaluation judge. To evaluate the effectiveness of LLMs as both text extractors and judges, we conducted a human annotation study with 200 samples and only three annotators. 

A key threat to validity is contamination bias, when evaluation examples (or close paraphrases) appear in a model’s training data, artificially inflating performance. To guard against this, we tested whether GPT-4o mini had been trained on our NeurIPS 2021–2022 dataset by providing only each paper’s title and prompting it to summarize the content and identify limitations. In every case, the model replied with a disclaimer indicating unfamiliarity with the specific work (e.g., “I am not familiar with the specific paper titled …”). This consistent outcome suggests the model lacked prior exposure to the full texts, supporting the integrity of our evaluation.

While we selected GPT-4o mini for text extraction, generation, and evaluation due to its superior performance, relying on a single LLM for these roles introduces several potential biases. We took specific steps to mitigate these risks:
To counter self-validation bias, where the model might favor its own output, we cross-referenced its judgments with human evaluations and incorporated RAG. For positional bias, where the model may favor the first input when comparing texts, we swapped the input order to ensure consistent results. To reduce confirmation bias, the tendency to generate generic limitations, we used RAG to introduce more diverse evidence. Finally, to check for hallucinations, three human annotators verified that all extracted limitations were grounded in the source text.
Although these strategies are crucial for improving reliability, we acknowledge that they do not completely eliminate these inherent biases.

For future work, we will expand our dataset to more diverse domains (e.g., bioinformatics, cognitive science) to test the cross-domain robustness of our models. We also plan to enhance our generation framework by exploring more advanced multi-agent and open-source LLMs via RAG. Finally, we will scale our human validation efforts with a larger, more diverse pool of expert annotators to enable a deeper and more reliable analysis. 



\section*{Ethics Statement}
This research adheres to ACL ethical standards. All data, including research papers and OpenReview feedback, were sourced from public repositories in compliance with their usage policies and were not filtered based on discriminatory attributes. Our user study involved three computer science graduate students who participated voluntarily with no conflicts of interest.

We acknowledge and address inherent LLM risks, including biases from training corpora, confirmation bias toward “safe” limitations, fluency and verbosity biases favoring longer or well-written outputs, and self-validation bias when using the same model for multiple tasks. To mitigate these, we (1) ground all generations in source content and peer reviews via a RAG framework to improve factuality and reduce verbosity; (2) diversify our ground truth by incorporating human-authored OpenReview critiques; (3) use multiple models to break self-validation circularity; and (4) conduct parallel human evaluations to detect overconfidence and other model-specific biases. We recognize that further work is needed to rigorously quantify these issues and plan to investigate cross-domain robustness in future studies.

\bibliography{anthology,custom}

\appendix

\section{Appendix} 

\label{sec:appendix}

In our PointWise evaluation method, we measured precision, recall, and F1 score from True Positive, False Positive, and False Negative. 
\subsection{Coverage Measurement}
We compute:
\[
\text{P}_{{r}_{i}} \;=\; \frac{\text{TP}_{i}}{\text{TP}_{i} + \text{FP}_{i}}, 
\quad
\text{R}_{{r}_{i}} \;=\; \frac{\text{TP}_{i}}{\text{TP}_{i} + \text{FN}_{i}},
\]
and the F\textsubscript{1} score is the harmonic mean of \(\text{P}_{{r}_{i}}\) and \(\text{R}_{{r}_{i}}\).

\subsection{Performance Measurement}
\label{performance-measurement}
\paragraph{A. Text-Based Evaluation.}
We apply standard text similarity metrics to each matched pair, including ROUGE-1, ROUGE-L, BERTScore, Cosine Similarity, Jaccard Similarity, and BLEU, calculating the number of overlapping unigrams, the longest sequence of words, and the similarity between contextual embeddings. 

\paragraph{B. Keyword-Based Evaluation.}
We employ KeyBERT \cite{grootendorst2020keybert} to extract a set of top keywords from the ground truth limitations \(K_{G_i}\) and from the LLM-generated limitations \(K_{H_i}\). We then measure the cosine and Jaccard similarity between \(K_{G_i}\) and \(K_{H_i}\) for each paper \(P_i\) and average these scores across the dataset.

\paragraph{C. Heading-Based Evaluation.}
We also compare concise ``headings'' or short titles for each limitation. Let \(T_{G_i}\) be the heading for \(G_i\) and \(T_{H_i}\) the heading for \(H_i\). We compute BERTScore between \(T_{G_i}\) and \(T_{H_i}\) for every paper \(P_i\) and then average these values. This provides a high-level measure of how closely the top-level concepts align.

By combining coverage and performance metrics in a PointWise manner, our framework provides a detailed assessment of how well an LLM-generated set of limitations captures the breadth and depth of the ground truth. This approach also facilitates fine-grained error analysis by examining matched pairs on a per-limitation basis.

We measure coverage for both ground truth and LLM-generated limitations \emph{independently}, focusing on each unique limitation within the matched pairs.

Furthermore, we conduct experiments using:
\begin{enumerate}
    \item The top three sections (\emph{Abstract, Introduction, and Conclusion})
    \item The entire paper (\emph{full paper})
\end{enumerate}
This setup enables us to examine how restricting the analysis to specific sections affects coverage and matching performance.

We used three distinct prompts to check the topic-level similarity between ground truth limitations and LLM-generated limitations (Figure \ref{fig:llm_as_a_judge_for_sub_limitations}, Appendix). To overcome the position bias, we choose the consistent one.

\begin{table*}[htbp]
\centering
\footnotesize
\begin{tabular}{cccccc}
\toprule
            & GPT-4 & Llama & HE1  & HE2  & HE3  \\ \midrule
GPT-4o mini & 1           & 0.71      & 0.9  & 0.92 & 0.95 \\
Llama   & -           & 1         & 0.81 & 0.79 & 0.76 \\
HE1         & -           & -         & 1    & 0.98 & 0.95 \\
HE2         & -           & -         & -    & 1    & 0.97 \\
HE3         & -           & -         & -    & -    & 1    \\ \bottomrule
\end{tabular}
\caption{Evaluating how good LLM `as a Judge' by checking Human Expert (HE) and model (GPT-4o mini, Llama-3.1 400B) agreement in determining whether an extracted limitation \textit{matches} a generated one (in PointWise Evaluation).}
\label{table:human-v-model-matching}
\end{table*}

\begin{table*}[htbp]
  \centering
  \footnotesize
  \begin{tabular}{l|c}
    \hline
    \textbf{Section} & \textbf{Cosine Similarity} \\ \hline
    \textbf{Abstract vs Limitation} & \textbf{33.27}  \\ 
    \textbf{Introduction vs Limitation} & \textbf{33.06} \\ 
    Related Work vs Limitation & 25.10 \\ 
    Methodology vs Limitation & 26.58 \\ 
    Dataset vs Limitation & 25.59 \\ 
    \textbf{Conclusion vs Limitation} & \textbf{33.04} \\ 
    Experiment and Results vs Limitation & 31.73 \\ \hline
  \end{tabular}
  \caption{Cosine Similarity between each section and the Limitation section. }
  \label{tab:cosine-sim}
\end{table*}

\begin{figure*}[htbp]
    \centering
    \begin{tikzpicture}
        \node[draw, rectangle, rounded corners, fill=yellow!20, text width=0.8\textwidth, inner sep=5pt, scale=1.0] (box) {
            \begin{minipage}{\textwidth}
                \small
                \textbf{Prompt} = \texttt{\textquotesingle\textquotesingle\textquotesingle} \\
                Here is the text containing extracted limitations. Please identify and list each limitation, ensuring that each one addresses a distinct topic or point.
                \texttt{\textquotesingle\textquotesingle\textquotesingle}
                                
            \end{minipage}
        };
    \end{tikzpicture}
    \caption{Prompt to extract limitations from ground truth text.}    \label{fig:prompt_topic_generation_related_text}
\end{figure*}

                                

\begin{figure*}[htbp]
    \centering
    \begin{tikzpicture}
        \node[draw, rectangle, rounded corners, fill=yellow!20, text width=0.8\textwidth, inner sep=5pt] (box) {
            \begin{minipage}{\textwidth}
                \small
                \raggedright 
                \textbf{Prompt} = \texttt{\textquotesingle\textquotesingle\textquotesingle} \\
                You are a helpful, respectful, and honest assistant for generating limitations or shortcomings of a research paper.
                I am providing 'Abstract', 'Introduction', 'Related Work', 'Methodology', 'Experiment and Results', 'Conclusion', and other sections of a scientific paper alongside the related cited papers texts. 
                Generate limitations based on these texts.
                \texttt{\textquotesingle\textquotesingle\textquotesingle}
            \end{minipage}
        };
    \end{tikzpicture}
    \caption{Prompt to generate limitations from Input and cited papers text.}
    \label{fig:prompt_lim_gen}
\end{figure*}

\begin{figure*}[htbp]
    \centering
    \begin{tikzpicture}
        \node[draw, rectangle, rounded corners, fill=yellow!20, text width=0.8\textwidth, inner sep=5pt, scale=1.0] (box) {
            \begin{minipage}{\textwidth}
                \small
                \textbf{Prompt 1} = \texttt{\textquotesingle\textquotesingle\textquotesingle} \\
                A tuple contains (list1, list2). Check whether both 'list1' and 'list2' have similar topics or limitation. If both have similar topics or limitations you can say "Yes", otherwise "No".
                Your answer should be "Yes" or "No" with explanation. 
                \texttt{\textquotesingle\textquotesingle\textquotesingle}
                
                \vspace{2mm}\hrule\vspace{2mm}
                \textbf{Prompt 2} = \texttt{\textquotesingle\textquotesingle\textquotesingle} \\
                A tuple contains (list2, list1). Check whether both 'list2' and 'list1' have similar topics or limitation. If both have similar topics or limitations you can say "Yes", otherwise "No".
                Your answer should be "Yes" or "No" with explanation.

                \vspace{2mm}\hrule\vspace{2mm}
                \textbf{Prompt 3} = \texttt{\textquotesingle\textquotesingle\textquotesingle} \\
                Check whether 'list2' contains a topic or limitation from 'list1' or 'list1' contains a topic or limitation from 'list2'. 
      Your answer should be "Yes" or "No" with explanation.
            \end{minipage}
        };
    \end{tikzpicture}
    \caption{LLM as a Judge for each limitation. We use three distinct prompts to verify consistency.
    }    \label{fig:llm_as_a_judge_for_sub_limitations}
\end{figure*}

\begin{table*}[htbp]
  \centering
  \small
  \begin{tabular}{p{5.4cm} p{5.4cm} c c c} 
    \toprule
    \textbf{Ground Truth} & \textbf{Gen. Lim.} & \textbf{GPT 4} & \textbf{Llama 3} & \textbf{User} \\ 
    \midrule
    \textbf{Model Complexity Concerns}: - There is a question regarding whether the performance gains are due to increased model complexity rather than the proposed recursive mixing approach. 
    & 
    \textbf{Potential for Increased Complexity}: Although the method is described as simple, the recursive nature of the approach may introduce complexity in implementation and understanding, particularly for practitioners who may not be familiar with the underlying concepts 
    & Yes & Yes & Yes \\ 
    \midrule
    \textbf{Computational Complexity}: The Dual-aspect Attention mechanism significantly increases computational complexity, which may affect the performance and practicality of the proposed method. 
    & 
    '**Complexity of the Model**: The DACT arch, while innovative, introduces additional complexity compared to traditional models', 'This complexity may lead to longer training times and increased computational resource requirements, which could be a barrier for practical applications in resource-constrained environments' 
    & Yes & Yes & Yes \\ 
    \midrule 
    \textbf{Insufficient Detailed Experimental Analysis}: The paper primarily provides quantitative results without a detailed experimental analysis, which limits the understanding of the findings. 
    & 
    \textbf{Limited Discussion on Failure Cases}: While the paper discusses success and failure cases of existing methods, it may not provide sufficient insight into the specific conditions under which SCILL might fail, limiting the understanding of its robustness. 
    & Yes & No & Yes \\ 
    \midrule 
    \textbf{Connection to Practical Methods}: There is an unknown connection between the theoretical findings and popular ensemble methods used in practice, which raises questions about the practical relevance of the work. 
    & 
    \textbf{Unexplored Variants of Ensemble Methods}: The research does not explore other ensemble methods beyond those mentioned, such as boosting or stacking, which may have different theoretical properties and could provide additional insights into the interpolating regime 
    & No & Yes & Yes \\ 
    \midrule 
    \textbf{Lack of Exploration of Alternative Methods}: The paper does not sufficiently explore or compare the performance of other methods like MFVI and SGLD in the context of covariate shift, which could provide a more comprehensive understanding of the problem. 
    & 
    \textbf{Absence of Dataset Details}: The dataset section is marked as "nan," indicating a lack of information about the datasets used for experimentation', 'This omission makes it difficult to assess the validity and applicability of the findings, as the choice of dataset can significantly influence the results 
    & No & No & No \\ 
    \bottomrule
  \end{tabular}
  \caption{Examples of Annotator, GPT 4o-mini, and LLama judgement on whether a generated limitation should be matched with a ground-truth limitation or not.}
  \label{tab:paper_gt_vs_gen_lim_vs_human}
\end{table*}

\begin{figure*}[htbp]
    \centering
    \begin{tikzpicture}
        \node[draw, rectangle, rounded corners, fill=yellow!20, text width=0.9\textwidth, inner sep=5pt, scale=1.0] (box) {
            \begin{minipage}{\textwidth}
                \small
                \textbf{Ground Truth Limitations:} Our goal is to provide comprehensive study of learning algorithms, including their hyperparameters. However, our study is limited on the variety of other design choices to restrict the experimental complexity. While we studied both synthetic and realistic image datasets, both these datasets are relatively simple with the same small number of generative factors and each of the factor follows uniform distribution. For learning algorithms, we focus on studying the inductive bias on the representation format while fixing the model arch design which can impact the results. Moreover, we did not study hyperparameters beyond those related to the latent representations. Specifically, we did not study how the type and configurations of the optimizer and the batch size would change the results; instead, we followed common setups in previous studies.

                \texttt{\textquotesingle\textquotesingle\textquotesingle}
                \vspace{2mm}\hrule\vspace{2mm}
                
                \textbf{LLM Extracted Limitations:} \\
                1. Here are the limitations extracted from the provided limitation:, 
                
                2. **Experimental Complexity**: The study is limited in the variety of design choices to reduce experimental complexity., 
                
                3. **Dataset Complexity**: The study utilizes relatively simple synthetic and realistic image datasets, which have a small number of generative factors., 
                
                4. **Generative Factors**: Each generative factor in the datasets follows a uniform distribution, limiting the diversity of the data., 
                
                5. **Model arch**: The focus is on studying the inductive bias related to the representation format while keeping the model arch design fixed, which may influence the results., 
                
                6. **Hyperparameter Scope**: The study does not explore hyperparameters beyond those related to latent representations., 
                
                7. **Optimizer and Batch Size**: The impact of different types and configurations of optimizers and batch sizes on the results is not examined, as the study adheres to common setups from previous research.
              
            \end{minipage}
        };
    \end{tikzpicture}
    \caption{Ground Truth Limitations and LLM Extracted Limitations in NeurIPS dataset.}    \label{fig:llm_refined_limitations_in_neurips}
\end{figure*}

\begin{figure*}[htbp]
    \centering
    \begin{tikzpicture}
        \node[draw, rectangle, rounded corners, fill=yellow!20, text width=0.9\textwidth, inner sep=5pt, scale=1.0] (box) {
            \begin{minipage}{\textwidth}
                \small
                 
                \textbf{Ground Truth Limitations:} While our work provides useful starting point for understanding student feedback,there are limitations to our work. Addressing these limitations will be an important area for future research. Comments may not reflect real student feedback. The comments in our dataset are from users who have chosen to post publicly on youtube. Addi- tionally,the comments may include features specific to this online education setting. Thus,the comments may reflect real student comments from these courses. There is selection bias in lecture sources. Sight includes lectures that may be drawn from the most successful offerings of that course. The instructional quality may not be representative of typical instruction. Thus,inferences drawn about the instruction should be interpreted with caution,as they might not generalize to other lecture settings. We analyze only english comments. We analyze only english comments because the lecture content is given in english and the authors are most comfortable with english. As result,our rubric may not capture the types of feedback from nonenglish students watching lectures taught in english. we annotate small subsample of the data to assess the validity of the automatic labels,we conduct diagnostic study on small,randomly selected subset of the dataset,comprising approximately of the comments. Our work aims to establish preliminary evaluation of the humanmodel agreement and model annotations,and further validation of the automatic labels is necessary. 

                \texttt{\textquotesingle\textquotesingle\textquotesingle}
                \vspace{2mm}\hrule\vspace{2mm}
            
                \textbf{LLM Extracted Limitations:} Here are the limitations extracted from the provided limitation: \\

            1. **Comments May Not Reflect Real Student Feedback**: The dataset consists of comments from users who have chosen to post publicly on YouTube, which may not accurately represent genuine student feedback.
            
            2. **Features Specific to Online Education Setting**: The comments may include characteristics that are unique to the online education context, potentially skewing the interpretation of student feedback.
            
            3. **Selection Bias in Lecture Sources**: The lectures included in the analysis may be drawn from the most successful offerings of a course, leading to a selection bias that does not represent typical instructional quality.
            
            4. **Generalizability of Instructional Quality**: Inferences about instructional quality based on the selected lectures should be made with caution, as they may not generalize to other lecture settings.
            
            5. **Analysis Limited to English Comments**: The analysis is restricted to comments in English, which may overlook feedback from non-English speaking students who are watching lectures taught in English.
            
            6. **Small Subsample for Validity Assessment**: The study annotates a small subsample of the data to assess the validity of automatic labels, which may not provide a comprehensive evaluation of the entire dataset.
            
            7. **Need for Further Validation of Automatic Labels**: The work establishes a preliminary evaluation of human-model agreement and model annotations, indicating that further validation of the automatic labels is necessary.
            \end{minipage}
        };
    \end{tikzpicture}
    \caption{Ground Truth Limitations and LLM Extracted Limitations in ACL dataset.}    \label{fig:llm_refined_limitations_in_neurips1}
\end{figure*}

\begin{figure*}[htbp]
    \centering
    \begin{tikzpicture}
        \node[draw, rectangle, rounded corners, fill=yellow!20, text width=0.9\textwidth, inner sep=5pt, scale=1.0] (box) {
            \begin{minipage}{\textwidth}
                \small
                \textbf{OpenReview:} \\
                \textbf{Reviewer 1:} This paper tries to tackle the domain generalization problem by finding flat minima. It is interesting to investigate the loss landscape of the DNNs on domain generalization problems. Originality: The method proposed in this paper is mainly built on previous approaches such as SWA, and the authors mention this in the submission. Quality: This submission is complete piece of work and the claims are well supported by the empirical results. The evaluation is reasonable. Clarity: This submission is well-written and easy to follow. Significance: This paper provides interesting empirical observations and improves the state-of-the-art by applying their proposed new approach. Pros: The proposed method consistently improves the model performance on domain generalization problems, and the proposed method is flexible and can be easily integrated with other training approaches for better performance. Cons: The generalization bound does not provide intuition for robust risk minimization.. \\
                
                \textbf{Reviewer 2:} Originality The paper is, perhaps by design, not particularly original. Rather, it makes the observation that an existing set of techniques with particular goal also has applications to different goal. This is still, to the best of my knowledge, novel observation and contribution, and furthermore the modifications made to SWA to arrive at SWAD also appear novel. Nevertheless, these modifications are relatively small, so the method can still be viewed as version of the general idea of SWA, and thus originality is not the paper' strong suit. Quality The paper is of relatively high quality, primarily due to the empirical results. did not carefully check the theory for correctness, but am unsure as to the significance of the theorems. They appear rather similar to prior theoretical results on how flat minima can lead to better generalization, but with some additional divergence terms thrown in as we are dealing with the DG setting... \\
                
                \textbf{Reviewer 3:} Pros: The paper tackles an important and relevant problem to the ML community. The proposed algorithm is relatively simple to implement in practice and perform well experimentally. Cons: My major concern is wrt to the novelty of the work and the connection/motivation between the proposed algorithm SWAD, domain generalization and the derived theoretical insights...

                \texttt{\textquotesingle\textquotesingle\textquotesingle}
                \vspace{2mm}\hrule\vspace{2mm}
                
                \textbf{LLM extracted OpenReview:} \\
                "1. . **Connection Between Theory and Method**: The relationship between the theoretical framework presented in the paper and the SWAD algorithm is weak. There is a need for a clearer connection between the flatness term in the equations and the algorithm's performance.", 
                
                '2. . **Novelty of Theoretical Results**: The novelty of the theorem compared to classic results, such as those from Ben-David, is questioned. The theoretical contributions do not significantly advance the understanding of flat minima in the context of domain generalization.', 
                
                '3. . **Generalization Bound Intuition**: The generalization bound provided does not offer sufficient intuition for robust risk minimization (RRM), and it does not convincingly demonstrate that RRM is superior to standard empirical risk minimization (ERM).'], 
                
                '4. . **Dependence on Previous Work**: The proposed method heavily relies on the existing SWA approach, with only minor modifications. This raises concerns about the originality of the contribution.', 
                
                '5. . **Theoretical Framework Limitations**: The reliance on VC dimension in the theoretical framework is problematic, as it can be excessively large for the models used, especially in domain generalization scenarios.', 
                
                '6. . **Overlooked Details in Proofs**: There are concerns that important terms have been overlooked in the technical proofs, which could undermine the validity of the results.', 
                
                '7. . **Lack of Clear Motivation for SWAD**: The motivation for using SWAD over SWA is not clearly articulated, making it difficult to understand the advantages of the proposed algorithm.', 
                
                '8. . **Introduction of New Hyperparameters**: The proposed algorithm introduces additional hyperparameters, which may complicate the implementation and tuning process.', 
                
                '9. . **Limited Societal Impact Discussion**: While a limitations section is included, there is a lack of discussion regarding the societal impacts of the research, although specific concerns were not identified.', 
                
                "10. . **Potential for Broader Applications**: The method's applicability beyond domain generalization is not explored, which could limit its significance and impact in the broader machine learning community.
              
            \end{minipage}
        };
    \end{tikzpicture}
    \caption{Tool extracted OpenReview and LLM Extracted OpenReview.}    \label{fig:llm_refined_limitations_in_neurips3}
\end{figure*}

\end{document}